\documentclass[prl,aps,twocolumn,showpacs]{revtex4}

\usepackage{graphicx}

\usepackage{graphicx}

\begin{document}

  \title{Enhanced Ferromagnetic Stability in Cu Doped Passivated GaN Nanowires}

  \author{H. J. Xiang}
  
  \affiliation{National Renewable Energy Laboratory, Golden, Colorado 80401, USA}

  \author{Su-Huai Wei}

  \affiliation{National Renewable Energy Laboratory, Golden, Colorado
  80401, USA}

\date{\today}

\begin{abstract}
  Density functional calculations are performed to investigate the room
  temperature ferromagnetism in GaN:Cu nanowires (NWs). Our results indicate 
  that two Cu dopants are most stable when they are near each other. Compared to
  bulk GaN:Cu, we find that magnetization and ferromagnetism in Cu doped NWs is
  strongly enhanced because the band width of the Cu $t_d$ band is reduced due to the 
  1D nature of the NW. 
  The surface passivation is shown to be crucial to sustain the ferromagnetism in GaN:Cu NWs. 
  These findings are in good agreement with experimental observations and indicate
  that ferromagnetism in this type of systems can
  be tuned by controlling the size or shape of the host materials.
\end{abstract}

\maketitle

Dilute magnetic semiconductors (DMSs) have attracted wide interest
recently because they possess both semiconducting and magnetic properties
at the same time, thus suitable for spintronic applications. One of the
major focus in this important research field is to produce DMSs with Curie temperatures 
($T_c$) at or above room temperature. It has been predicted that the use of 
wide band-gap semiconductors, e.g., GaN, as the host material for DMSs can
lead to high $T_c$ \cite{Dietl2000}. However, experimental studies on 
transition-metal (TM)-doped bulk GaN have led to conflicting results.
For example, some experiments have shown the existence of
ferromagnetism in Ga$_{1-x}$Mn$_x$N but with $T_c$
varying between 8 K \cite{Sarigiannidou2006} and 904 K \cite{Sonoda2002} at 
$x=0.06$. In contrast, a magneto-optical study \cite{Ando2003} suggested 
that Ga$_{1-x}$Mn$_x$N is paramagnetic. Theoretical studies also
show that the magnetic properties of bulk Ga$_{1-x}$Mn$_x$N is quite unique:
depending on the Mn concentration, carrier density, and pressure, it can change
from ferromagnetism to antiferromagnetism \cite{Dalpian2005,Dalpian2006}.

Recently, Wu {\it et al.} \cite{Wu2006} predicted that two
Cu ions in a configuration about 6.2 \AA\ from each other
in bulk GaN are coupled ferromagnetically with the total moment
2 $\mu_B$/Cu. Subsequently, room temperature ferromagnetism was found
in Cu-implanted GaN samples, although with a much smaller saturation 
magnetization (from 0.01 to 0.27  $\mu_B$/Cu), and the results are very
sensitive to the annealing temperature \cite{Lee2007,Seipel2007}. 
A more recent calculation by Rosa and Ahuja \cite{Rosa2007} points out that the
ferromagnetic coupling in bulk GaN:Cu is much weak than expected from
previous calculation \cite{Wu2006} because the magnetic moment on Cu
is very sensitive to the Cu-Cu distance; it becomes much smaller
when Cu-Cu becomes nearest neighbor. However, a recent experimental study
by Seong {\it et al.} reported that room temperature ferromagnetism can be
achieved in Ga$_{1-x}$Cu$_x$N nanowires (NWs) 
synthesized in a chemical vapor transport system under flow of NH$_3$ 
and the saturation magnetic moment is significantly higher than that in 
bulk GaN ($\sim$1.00 $\mu_B$/Cu at 5 K) \cite{Seong2007}. 
The dramatically different saturation magnetic moment of Cu in bulk GaN and GaN
NWs suggest that the magnetic behavior of Cu doped GaN could be manipulated
by tuning the size or shape of host. However, the mechanism of the enhanced
magnetic coupling in GaN:Cu NWs is not known, including what is the
effect of the surface passivation in the NWs.    

In this Letter, we perform a comprehensive first principles study
to understand the structural and magnetic properties of Cu doping 
in bulk GaN and GaN NWs. Both bare and passivated
GaN NWs are studied. Our first-principles spin-polarized density functional
theory (DFT) calculations were performed on the basis of the projector
augmented wave method \cite{PAW} encoded 
in the Vienna ab initio simulation package \cite{VASP}
using the generalized-gradient approximation (GGA) 
\cite{Perdew1996} and the plane-wave
cutoff energy of 400 eV. For relaxed
structures, the atomic forces are less than 0.02 eV/\AA.

\begin{figure}
  \begin{center}
  \includegraphics[width=7.0cm]{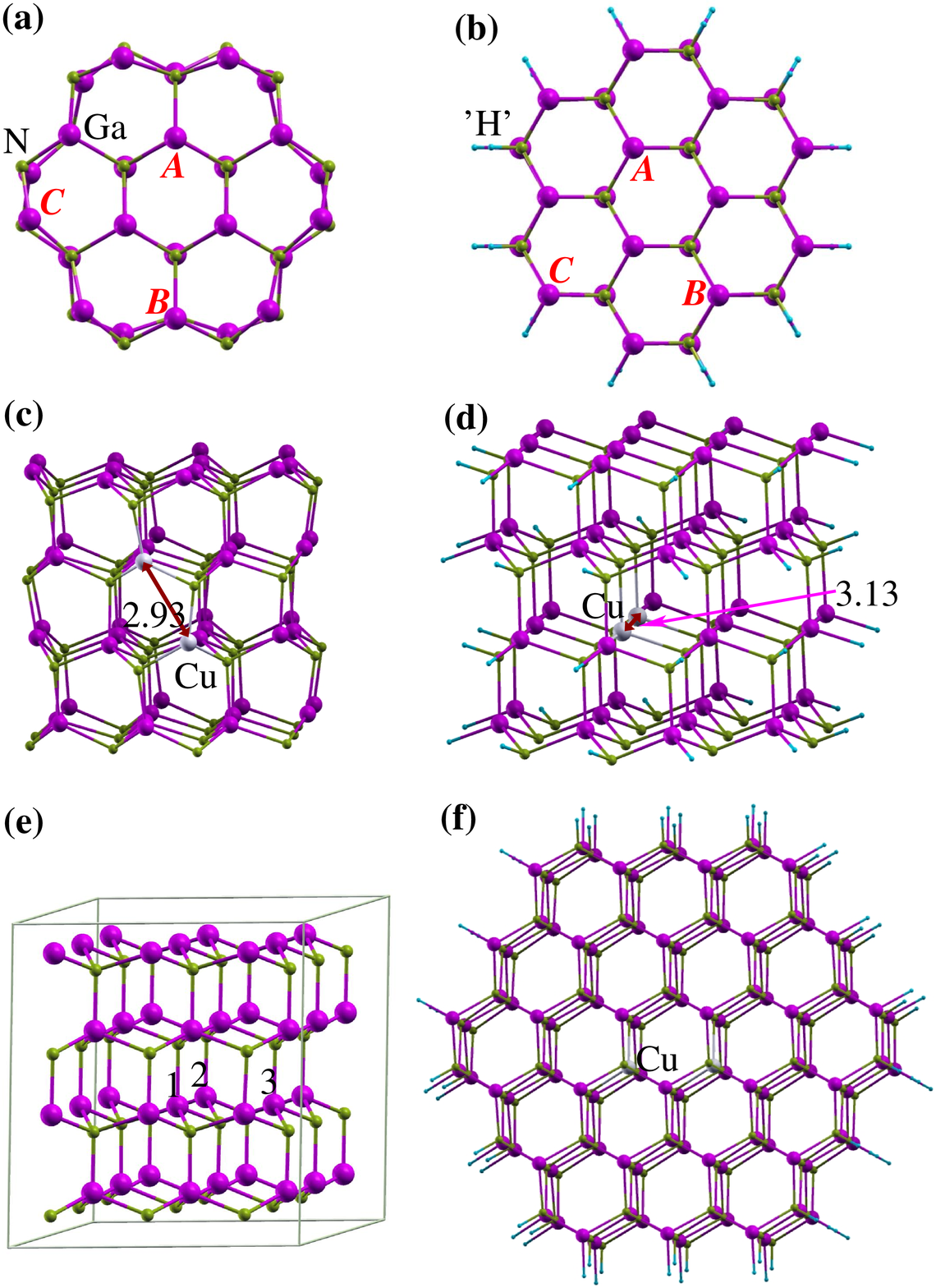}
  \caption{(a) Top view of a bare GaN NW with a diameter about 1.0 nm,  (b) top view
  of a GaN NW passivated by pseudo H atoms, (c) side
  view of the most stable configuration of two substitutional Cu 
  atoms in the bare GaN NW, (d) side
  view of the most stable configuration of two substitutional Cu atoms
  in a passivated GaN NW, (e) side view of a $3\times 3
  \times 2$ bulk GaN supercell, and (f) top view of a
  configuration of two substitutional Cu atoms
  in a passivated GaN NW with a larger diameter 1.6 nm. A, B, and C in (a) and
  (b) denote three inequivalent Ga positions. The numbers in (c) and
  (d) give the distance (in \AA) between two Cu dopants. Number 1, 2 ,
  and 3 in (e) label different Ga atoms in the supercell.} 
  \label{fig1}
  \end{center}
\end{figure}

For bulk GaN, the optimized wurtzite lattice
constants are a$=3.218$ \AA\ and c$=5.240$ \AA\ with the internal
parameter u=$0.3767$ (experimental values: a$=3.189$ \AA, c$=5.186$
\AA, and u=$0.377$). The calculated band gap for bulk GaN is 1.72 eV.
The GaN NWs are orientated along the [0001] direction with a diameter
about 1 nm, as shown in Fig.~\ref{fig1}(a) and (b). For passivated
NWs, pseudo-hydrogen is used to  saturate the dangling bond of the NW surface.
In the simulation, the NW axis is along the $c$ direction, and the
lateral supercell size is chosen so that the closest distance between
two neighbor NWs is larger than 8 \AA. After optimization, the diameters of the GaN NWs 
become slightly smaller, and consequently, to minimize the
strain, the lattice constant $c$ is increased by 0.018 \AA\ and 0.046 \AA\ for
passivated and bare GaN NWs, respectively, as in the case of bare ZnO NWs
\cite{Xiang2006}. Both bare and passivated NWs are found to have a
direct bang gap (1.67 eV and 3.25 eV, respectively) at $\Gamma$. The small
band gap of the bare wire is due to the formation of surface defect levels.
In case of Cu doping in NWs, we use  a $1\times 1 \times 2$
supercell with a $1\times 1 \times 4$ k-mesh. 
For comparison, Cu doping in bulk GaN is also
studied by using a $3\times 3 \times 2$ supercell (shown in Fig.~\ref{fig1}(e)) with a 
$4\times 4 \times 4$ k-mesh. The lattice constants of the doped
systems are fixed to those of the undoped hosts since substitutional
Cu changes only marginally the lattice constant \cite{Wu2006,Seipel2007}.

To compare the dopability of Cu in bulk GaN and GaN NWs 
and identify stable dopant position, we first calculate the 
formation energy of Cu substitution for Ga in GaN host as 
\begin{equation}
  \label{eq1}
\begin{array}{ccl}
  \Delta H_f &=&
  E(\mathrm{GaN:Cu})-E(\mathrm{host})+\mu(\mathrm{Ga})-\mu(\mathrm{Cu})
  \\
  &=&\Delta E+\mu(\mathrm{Ga})-\mu(\mathrm{Cu})
\end{array}
\end{equation}
where $E(\mathrm{GaN:Cu})$ is the total energy of the doped system, $E(\mathrm{host})$
denotes the  total energy of the GaN host for the same supercell
in the absence of the defect, $\mu(\mathrm{Ga})$ and $\mu(\mathrm{Cu})$ are the chemical
potential for Ga and Cu, respectively. We note that the absolute
formation energy depends on $\mu(\mathrm{Ga})$ and
$\mu(\mathrm{Cu})$. However, it is sufficient to calculate $\Delta E$
for the purpose of comparing the relative stability of Cu in GaN host.
For a NW as shown in Fig.~\ref{fig1}(a) and (b), there are three
inequivalent Ga positions, i.e., A, B, and C. In all cases, we confirm that Cu
will not form AX center in GaN due to the delocalization of the Cu
3d related defect bands. Our results are reported in Table.~\ref{table1}.  
For bare GaN NW, we find that the formation energy of Cu$_{\mathrm{Ga}}$ 
is smaller than that in bulk GaN, because Cu$_{\mathrm{Ga}}$ can relax
more easily in the NW than in the bulk. The lowering of the formation
energy is most dramatic at the surface of the NW, where the formation energy
is about 0.96 eV smaller. 
For the passivated GaN NW, we find that the formation energy of
Cu$_{\mathrm{Ga}}$ is about 0.3 eV larger than that in bulk GaN and is most stable
at the A site. This result follow the general trends observed in
nanocrystal quantum dots \cite{Li2008} 
suggesting that the increased formation energy for this non-isovalent dopants
is mainly due to the quantum confinement-induced band gap increase, but not
sensitive to the shape of the quantum structure. 
The above results show that the doping ability of
Cu in GaN NW depends sensitively on the surface termination of the NW. 
We also find that in all cases with an isolated Cu atom in GaN
hosts, the total magnetic moment is 2 $\mu_B$ and the local moment on
Cu is about 0.65 $\mu_B$ due to the strong Cu 3d and N 2p hybridization \cite{Wu2006}.

Experimentally, the valence state of Cu in Cu doped
GaN NW is found to be close to that of Cu in CuO \cite{Seong2007}.   
To confirm this theoretically, we calculate the charge tranfer in Cu
doped NWs using atoms in molecules (AIM) theory \cite{Bader}.
We find that the valence state of Cu dopant in bare (passivated) GaN
NW is about 0.84 (0.88). 
For CuO and Cu$_2$O, the calculated Cu valence state is 0.99 and 0.55, respectively.  
So, the valence state of Cu dopant in GaN NW is indeed close
to that in CuO. The calculated valence state is not close to
the nominal value due to the strong covalency in the system.
The results also indicate that Cu occupies Ga site in the experimental sample \cite{Seong2007}.  

To investigate the interaction between Cu dopants in GaN NWs, we
substitute two Ga atoms with two Cu atoms. The most stable
configuration of two Cu atoms in GaN NWs is shown in
Fig.~\ref{fig1}. We can see that two Cu atoms tend to be next to each
other, i.e., they are bonded to the same N atom. This is due to the
strong bonding interaction between Cu 3d $t$ states.
In case of passivated
NW, both Cu atoms occupied A positions. In contrast, these Cu
atoms occupied C positions for the bare GaN NW case. This is
understandable since an isolated Cu atom prefers to occupy the C (A)
position in case of bare (passivated) GaN NW. 

\begin{figure}
  \begin{center}
  \includegraphics[width=8.5cm]{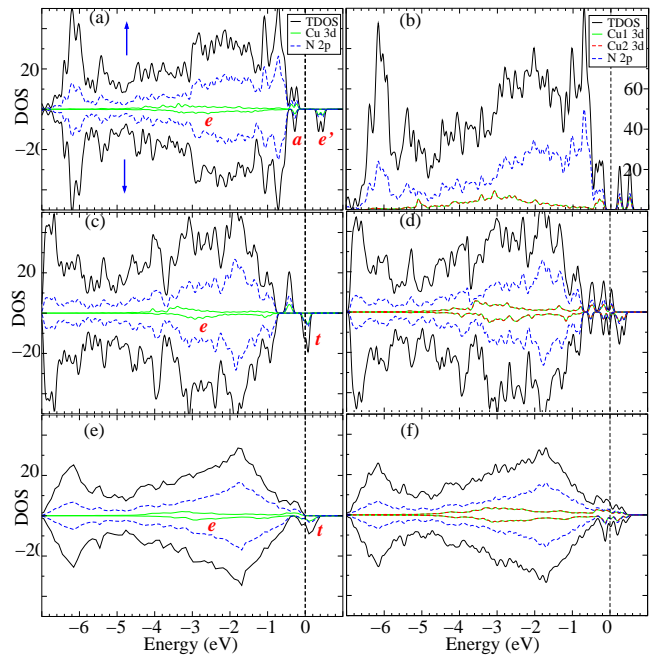}
  \caption{DOS of (a) a bare GaN NW with an
    isolated Cu dopant at the C position (Fig.~\ref{fig1}(a)), (b) a
    bare GaN NW with two Cu dopants (Fig.~\ref{fig1}(c)), (c)
    a passivated GaN NW with an isolated Cu dopant at the A
    position (Fig.~\ref{fig1}(b)), (d) a passivated
    GaN NW with two Cu dopants (Fig.~\ref{fig1}(d)), (e) an isolated
    Cu dopant in a bulk GaN supercell (Fig.~\ref{fig1}(e)), and (f) 
    two Cu dopants at positions 1 and 2 in a bulk GaN supercell
    (Fig.~\ref{fig1}(e)),
    respectively. Here, high energy conduction bands of
    GaN hosts are not shown.}  
  \label{fig2}
  \end{center}
\end{figure}

The spin exchange interaction between Cu 3d moments in GaN NWs
is studied through calculating different spin states, i.e., ferromagnetic (FM)
and antiferromagnetic (AFM) states. In case of Cu doped passivated GaN NW
(Fig.~\ref{fig1}(d)), the FM state with a total moment 2.8 $\mu_B$ is more favorable over the AFM
state by 90 and 31 meV for the unrelaxed and relaxed structures,
respectively. Surprisingly, for the doped bare GaN NW as shown in 
Fig.~\ref{fig1}(c), it is a non-magnetic semiconductor with a gap of 0.44 eV
within the d bands. To probe the origin of the different nature of
magnetism in these two cases, we plot the density of states (DOS) in
Fig.~\ref{fig2}. To facilitate the analysis, the DOSs of the systems
with an isolated Cu dopant are also shown. 
In both cases, we can see that there is a band
gap in the spin up component. 
In case of the passivated NW, the spin down $t$ orbitals
(In the crystal field of a wurtzite system, 3d orbitals are
splitted into low-lying two-fold $e$ states and  three nearly
degenerated $t$ orbitals) are partially occupied. However, when a Cu atom
occupies the surface position [position C in Fig.~\ref{fig1}(a)] of
the bare GaN NW, the different crystal field arising from a missing neighbor N
atom at the surface will split the $t$ orbitals into a low-lying one-fold $a$ and a
high-lying two-fold $e'$ sets. The presence of the band gap is
consistent with the stability of the surface Cu atom. 
When the two Cu dopants are next to each other [Fig.~\ref{fig1}(c)], the  
coupling between two Cu $t$-derived orbitals will result in a splitting between
the bonding and anti-bonding orbitals. Since the distance between two Cu atoms in bare NW is
only 2.93 \AA\ (the Cu-Cu distance in bulk GaN and passivated GaN NW
are  3.13 \AA\
and 3.12 \AA\, respectively), the splitting will
be larger than the spin exchange splitting, resulting in the low spin
non-magnetic state. Similar results are observed in GaN:Mn, where
pressure or surface strain can turn the system from a high-spin to a
low-spin configuration, diminish the ferromagnetism \cite{Dalpian2005,Wang2004}.   
Our calculations reveal that passivation of the surface is crucial in
this system to obtain ferromagnetism for Cu doped GaN NWs. It is noted
that previous theoretical studies focused only on TM (Mn
\cite{Wang2005A} and Cr \cite{Wang2005B}) doped GaN {\it bare} NWs.
In addition, the coupling between a pair of Cr atoms substituted
in unsaturated GaN nanoholes was found to be FM \cite{Wang2007}.

 \begin{table}
   \begin{center}
   \caption{\label{table1}
   Relative formation energy $\Delta E$ of an isolated
   Cu$_{\mathrm{Ga}}$ defect in bare and passivated GaN
   NWs. Please refer to Fig.~\ref{fig1} for the definition of
   positions A, B, and C. $\Delta E$ is related to the absolute
   formation energy $\Delta H$ by Eq.~\ref{eq1}. For bulk GaN, $\Delta
   E$ of an isolated  Cu$_{\mathrm{Ga}}$ defect is 3.33 eV.} 
   \begin{tabular}{cccc}
     &site A & site B & site C \\
     \hline
     Bare NW& 3.05 &3.27&  2.37 \\
     Passivated NW&3.52 & 3.68 & 3.52\\
   \end{tabular}
   \end{center}   
 \end{table}

To compare the different magnetic behavior
of Cu doping in bulk GaN and GaN NW, we also study the Cu-Cu interaction
in bulk GaN. First, we find that Cu dopants in bulk GaN also tend to be
nearest neighbor. The most favorable configuration is that two Cu
atoms occupy the in-plane nearest neighbor positions [positions 1 and
2 in Fig.~\ref{fig1}(e)] with a magnetic moment of 0.6 $\mu_B$/Cu. 
However, the energy difference between the FM 
and the AFM state in this configuration is only $-1.6$ meV (the value
for the unrelaxed structure is 10.5 meV), which is 
much smaller than that in passivated NW and in the case when the Cu-Cu
distance is about 6.2 \AA\ \cite{Wu2006}. 
In some other configurations, 
we find that the AFM state could be even more stable than the
FM state. For instance, if two Cu atoms occupy
positions 1 and 3 [Fig.~\ref{fig1}(e)], the FM state
has higher energy by 41 meV. These results indicate that 
FM in Cu doped bulk GaN is not robust, which is consistent with experimental
observations and the recent calculations of Rosa and Ahuja \cite{Rosa2007}. 

\begin{figure}
  \begin{center}
  \includegraphics[width=7.0cm]{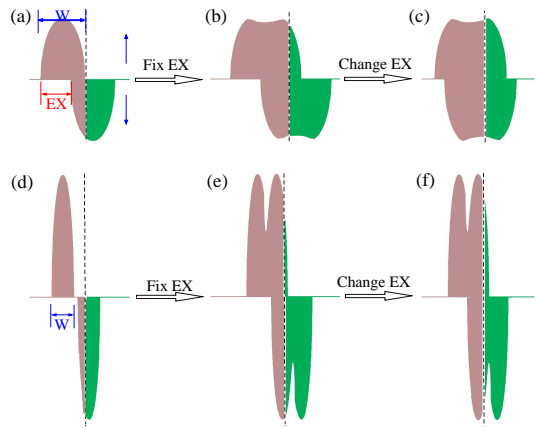}
  \caption{Schematic illustration of the origin of the different magnetic
  behavior between Cu doped passivated GaN NW and Cu doped bulk
  GaN.  (a) shows the DOS for the $t$ orbitals of an isolated Cu atom
  in bulk GaN, (b) and (c) are the DOSs of two
  in-plane neighbor Cu atoms in bulk GaN in the FM state with the
  exchange splitting (EX) fixed and varied, respectively. (d), (e),
  and (f) are the corresponding DOSs for Cu in a passivated GaN NW
  (Fig.~\ref{fig1}(d)). ``W'' denotes the band width.}
  \label{fig3}
  \end{center}
\end{figure}

To unravel the origin of the different behavior of Cu dopants in bulk
GaN and passivated GaN NWs, we illustrate the DOS of the $t$ orbitals
near the fermi level in Fig.~\ref{fig3}.
For the isolated Cu dopant, the main difference between bulk GaN:Cu and
GaN:Cu NW is that the band width of the $t$ states is smaller in NWs than that
in bulk GaN (0.13 eV v.s. 0.53 eV) due to the quasi-1D nature of
NWs. The DOSs for an isolated Cu in bulk GaN and GaN
NWs resemble those shown in Fig.~\ref{fig3}(a) and (d), respectively.
When two Cu dopants are close to each other, the $t$ orbitals in the
same spin will couple to each other, resulting in a low-lying bonding set and a high-lying
anti-bonding one in the FM state, broadening  the band. This stabilize the FM state over the AFM
state due to the more occupation of the spin-down bonding orbitals than the anti-bonding
state \cite{Dalpian2006,Sluiter2005}. However, it also increases the overlapping
between the spin-up and spin-down 
bands, causing electron transfer from the spin-up states to spin-down states as
shown in [Fig.~\ref{fig3}(b) and (e)], reducing the magnetic
moment. The transfer is large in bulk GaN 
than in GaN NW because the band width in bulk GaN is wider. The reduced magnetic moment
further leads to a reduced exchange splitting and more charge transfer. The final DOSs of 
the FM state for two Cu atoms in bulk GaN is shown in Fig.~\ref{fig3}(c). This explains why
Cu-Cu pairs in bulk GaN has a very small magnetic moment when the Cu-Cu distance is small
and why the magnetic interaction is weak in bulk GaN. This is because for bulk GaN:Cu, when 
Cu-Cu distance is small, the coupling between Cu $d$ orbital is large, but the magnetic
moment is small. When the Cu-Cu distance is large, the magnetic moment is recovered, but the 
$d$ orbital coupling becomes small.
On the other hand, the Cu $t$ orbital
band width is much smaller in GaN NW, thus the charge transfer and reduction of magnetic moment is
small in GaN:Cu NWs even when the Cu-Cu distance is small [Fig.~\ref{fig3}(f)]. 
This explains why magnetization and ferromagnetism is enhanced in GaN:Cu NWs.

\begin{figure}
  \begin{center}
  \includegraphics[width=6.0cm]{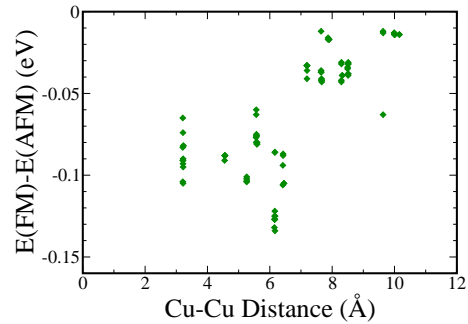}
  \caption{Energy difference between the AFM and FM state for all
  possible inequivalent configurations of two Cu dopants in a passivated GaN
  nanowire as shown in Fig.~\ref{fig1}(b).}
  \label{fig4}
  \end{center}
\end{figure}

Due to the configurational entropy, Cu dopants might have many
different configurations in GaN NWs under experimental growth
condition. Here to examine the dependence of the magnetic properties
on the configuration, we study all possible inequivalent
configurations (102 configurations in total) of two Cu
atoms in passivated GaN NW. To reduce the computational amount, no
structural relaxation is performed.  
It turns out that in all cases, FM
is always preferred over the AFM state, as shown in
Fig.~\ref{fig4}. Interestingly, the maximum magnitude of the energy
difference occur at some configurations with the Cu-Cu distance about 6
\AA, instead of the nearest neighbor configurations. This is because when
the Cu-Cu distance is smaller than 6 \AA\, the magnetic moment start to
decrease, thus reducing the magnetic interactions, as discussed above.
However, unlike in bulk GaN:Cu, the energy difference is still negative and
large at nearest neighbor Cu-Cu distance.
Thus FM in Cu doped passivated GaN NW is robust
with respect to Cu configurations in GaN NWs. 

In the experimental work by Seong {\it et al.} \cite{Seong2007}, the diameter of the
GaN NW is larger than that of the NW we discussed above.
Here to investigate the magnetic interaction between Cu dopants in
larger GaN NWs, we also consider two Cu dopants in passivated
GaN NW with a diameter about 1.6 nm, as shown in Fig.~\ref{fig1}(f).    
From our calculation, the FM state is more stable by 
75 and 54 meV than the AFM state for the unrelaxed and relaxed structures,
respectively. So the change of exchange interaction between Cu atoms in GaN
NW is not very significant when the diameter of the NW changes.

In summary, we explained why the ferromagnetism of GaN:Cu is
enhanced in passivated GaN NWs. We show that due to the 1D nature of
the NW, the band width of the Cu $t_d$ band is reduced, thus increases 
the magnetization. 
It is found that passivation of the nanowire surface has a significant
impact on the substitutional position, and thus on the magnetic
properties: For the bare GaN NWs, Cu tends to substitute Ga on the
surface, and the strong direct
interaction between two neighboring surface substitutional Cu atoms
results in a non-magnetic semiconducting state. However, when the surface is
passivated, Cu tends to substitute Ga 
inside the NW, and leads to ferromagnetism arising from the $d$-$d$ exchange
interactions. Our findings, thus, indicate that surface passivation
of the NW is crucial to sustain the FM state. 

This work is supported by the U.S. Department of
Energy, under Contract No. DE-AC36-99GO10337.
We thank Dr. Juarez L. F. Da Silva and Dr. Sukit Limpijumnong 
for useful discussions.


\end{document}